\RequirePackage{amsmath}
\documentclass[runningheads]{llncs}  

\usepackage{graphicx}
\usepackage{booktabs}
\usepackage{todonotes}
\usepackage{balance}

\usepackage{amssymb}

\newcommand{\farm}[0]{{\sc FARM}}
\newcommand{\dash}[0]{{\sc DASH}}

\newcommand{\ignore}[1]{}

\usepackage{tikz} 
\usepackage{xcolor}

\begin{document}


\title{Massive Multi-Agent Data-Driven Simulations of the GitHub Ecosystem} 



\author{Jim Blythe\inst{1} \and John Bollenbacher\inst{2} \and Di Huang\inst{1} \and Pik-Mai Hui\inst{2} \and Rachel Krohn\inst{3} \and Diogo Pacheco\inst{2} \and Goran Muric\inst{1} \and Anna Sapienza\inst{1} \and Alexey Tregubov\inst{1} \and Yong-Yeol Ahn\inst{2} \and Alessandro Flammini\inst{2} \and Kristina Lerman\inst{1} \and Filippo Menczer\inst{2} \and Tim Weninger\inst{3} \and Emilio Ferrara\inst{1}}  

\authorrunning{J Blythe \and J Bollenbacher \and D Huang \and PM Hui \and R Krohn \and D Pacheco \and G Muric \and A Sapienza \and A Tregubov \and YY Ahn \and A Flammini \and K Lerman \and F Menczer \and T Weninger \and E Ferrara}

\institute{USC Information Sciences Institute, Marina del Rey, CA (USA) \email{blythe|dihuang|gmuric|annas|tregubov|lerman|ferrarae@isi.edu}
\and Indiana University, Bloomington, IN (USA) \email{jmbollen|huip|pacheco|yyahn|aflammin|fil@iu.edu}
\and University of Notre Dame, 	Notre Dame, IN (USA) \email{rkrohn|tweninger@nd.edu}}

\maketitle

\begin{abstract}  
Simulating and predicting planetary-scale techno-social systems poses heavy computational and modeling challenges. The DARPA SocialSim program set the challenge to model the evolution of GitHub, a large collaborative software-development ecosystem, using massive multi-agent simulations. We describe our best performing models and our agent-based simulation framework, which we are currently extending to allow simulating  other  planetary-scale  techno-social  systems. The challenge problem measured participant's ability, given 30 months of meta-data on user activity on GitHub, to predict the next months' activity as measured by a broad range of metrics applied to ground truth, using agent-based simulation. The challenge required scaling to a simulation of roughly 3 million agents producing a combined 30 million actions, acting on 6 million repositories with commodity hardware. It was also important to use the data optimally to predict the agent's next moves. We describe the agent framework and the data analysis employed by one of the winning teams in the challenge. Six different agent models were tested based on a variety of machine learning and statistical methods. While no single method proved the most accurate on every metric, the broadly most successful  sampled from a stationary probability distribution of actions and repositories for each agent. Two reasons for the success of these agents were their use of a distinct characterization of each agent, and that GitHub users change their behavior relatively slowly.

\end{abstract}


\section{Introduction}


Two significant challenges on the way to realizing the promise of agent-based social simulation for policy evaluation and social science are making effective use of available data and scaling to planetary-sized cognitive agent simulations. As a first step, the DARPA SocialSim challenge problem measured participant's ability, given 30 months of meta-data on user activity on GitHub, to predict the next months' activity as measured by a broad range of metrics applied to ground truth, using agent-based simulation. The challenge 
involved making predictions about roughly 3 million individuals taking a combined 30 million actions on 6 million repositories.
We found that simulations of small subsamples of the population, on the order of tens or hundreds of thousands of agents, led to inconsistent and often misleading results, so a full-scale simulation was developed.
It was also important to use the data optimally to predict the agent's next moves. We describe the agent framework and the data analysis employed by one of the top performers in the challenge. The team used a variety of learning methods contributing to six different kinds of agents that were tested against a wide range of metrics. While no single method proved the most accurate on every evaluation metric, the broadly most successful of those tried sampled from a stationary probability distribution of actions and target repositories for each agent. Two reasons for the success of this agent were that individuals on GitHub change their behavior relatively slowly and that distinct information was maintained for each agent without generalizing across agents.
Our work that improves the performance achieved during the timeline of the challenge builds on these agents to incorporate novel behavior through further data analysis.

This paper makes the following contributions:
%
    First, we describe the agent-based simulator we developed to carry out massive-scale simulations of techno-social systems, which provides support for simulations of cognitive behavior and shared state across multiple compute nodes. 
    Second, we present the inference methods that we employed to implement different agent-based models, based on statistical modeling of historical activity, graph embedding to infer future interactions, Bayesian models to capture activity processes, and methods to predict the emergence of new users and repositories that did not exist in the historical data. These are novel applications of existing analytical tools to derive agent models from available data.
    Third, we provide a rigorous evaluation of the performance of six different models, as measured by a wide range of metrics, in simulating different scenarios concerned with user and repository popularity and evolution, as well as multi-resolution accuracy at the level of individual agents, groups of agents (e.g., communities, or teams), and the whole system. 
    We also describe the DARPA SocialSim GitHub Challenge, provide a  characterization of its rules, and describe how our team tackled it.
Our platform and models are general in scope, and have also been applied to large-scale agent simulations of behavior on the Twitter and Reddit social media platforms.

\section{Challenge problem description}

GitHub is a software social network where users interact with each other by contributing to repositories, following or becoming a member of specific projects, and interact with repositories through actions such as forking, committing, etc.
It is interesting to simulate since it combines aspects of social networks and collaborative work, and simulation may provide insights into team formation and productivity as well as aspects of widely-used code including the spread of vulnerabilities.
The DARPA SocialSim Challenge aims at simulating specific types of interactions between users and repositories on GitHub over time. In particular, it focuses on the simulation of social structure and temporal dynamics of the system, as well as looking at individual, community and population behaviors.  

There are ten event types in the model: create or delete either a repository, a tag, or a branch (respectively \textit{Create} and \textit{Delete}), create or comment a pull request (respectively \textit{PullRequest} and \textit{PullRequestReviewComment}), create an issue (\textit{Issues}, \textit{IssueComment}), and push (\textit{Push}, \textit{CommitComment}). Moreover, a user can \textit{watch} and \textit{fork}  existing repositories. Note that in the GitHub API a \textit{watch} event corresponds to starring a repository.

Given as an input the temporal information of users' actions on specific repositories, we aim at predicting future events of GitHub providing our simulation output in the following format: time, eventType, userID, repoID. Furthermore, the simulation aims at modeling not only the future events of existing users and repositories but also the creation/deletion of new repositories and users.

The training set we used for our simulation comprises all events of public users and repositories in the period spanning from 8/1/17 - 8/31/17 and 1/17/18 - 1/31/18, as well as metadata such as repository languages, user types etc. This includes a total of about $2.0M$ users and $3.3M$ repositories.
For the challenge, 
we were asked to simulate the events, users, and repositories of GitHub from 2/1/18 - 2/28/18. 
As the training set included a gap of $4.5$ months, 
additional information about the state of the system was provided, 
including all the profiles from users and repositories that were created during the gap. 

\section{Agent framework and domain implementation} 
In this section we describe the simulation framework and the model of GitHub that was shared by all agents developed for the challenge. We also describe steps 
to scale the simulation efficiently to millions of agents and repositories.

To implement our agent models we used \farm --- an agent-based simulation framework implemented in Python that supports large-scale distributed simulations \cite{blythe-tregubov18}. 
\farm\ also keeps track of the repeated and systematic experimentation required to validate the results from multi-agent simulations.
\farm\ supports agents developed with the \dash\ framework ~\cite{blythe12}, although it may be used with any agent through an API. The \dash\ agent framework supports simulations of cognitive behavior, and includes support for dual-process models, reactive planning and 
spreading activation.

In our experiments, \dash\ agents represent GitHub users and implement GitHub events. 
Agents in FARM can communicate either directly or by taking actions that are sent to a shared state object, called a hub, that can be observed by other agents. In the GitHub simulation model, every action taken by a user acts on a repository, so communication is modeled indirectly by sending actions to a hub that maintains the state of a set of repositories and provides information to agents about their repositories of interest. 

When all the agents and state relevant to a simulation reside in a single image, communication and action are efficiently implemented as method calls. However as we scale to millions of agents and repositories, some hardware cannot accommodate all the relevant state on a single host. \farm\ provides a multi-process infrastructure to divide agents and their state across multiple hosts as needed in a way that is transparent to the agent developer~\cite{blythe-tregubov18}. One hub is present on each image and shared state is managed with Apache ZooKeeper.
%
Frequently interacting users and repositories can be allocated on the same compute node to minimize cross-host communications. Using a multi-level graph partitioning algorithm to minimize the amount of communication across partitions based on the user-repository links found in the training data, simulation time was reduced by 67\% ~\cite{blythe-tregubov18}. 
We further improved performance with demand-driven shared state, so that repositories were only synchronized between compute nodes when agents on different nodes began to interact with them.
Finally, we modified \dash\ to reduce the space requirements of each agent so that the required simulations, with around 3 million agents taking a combined 30 million actions on 6 million repositories, run on a single host with 64GB of memory in around 20 minutes.

\section{Agent models} 

\subsection{Stationary probabilistic models}
\label{sec:stationary-prob-models}
A stationary probabilistic model is a simple way to describe user behavior with a finite number of available actions. In the following models, each user's actions are determined by a stationary probability distribution built from the past history of events the user has initiated. Two aspects of each user's behavior are determined: their overall event rate and the probability of each action. Both of these parameters are computed individually for each user. 

We implemented three variations of probabilistic simulation models. The first model selects an event type and independently selects the repository on which the selected action is to be applied. 
We refer to this as be {\em baseline model}.

The second model, called the {\em ground-event model}, selects an event type and repository simultaneously. As in the baseline model, in the ground-event model frequencies are computed from historical data but in this case the frequency is computed for each event and repository pair.

The third model, called the {\em preferential attachment model}, extends the baseline model by redefining user behaviour for \textit{watch} and  \textit{fork} events. When a user agent decides to watch a new repository, it first selects a neighboring user, who also worked on a repository this user interacted with, and then selects a repository with which the selected neighbor previously interacted. The neighboring user and repository are selected based on their popularity.

In all models, the frequency of users' actions is determined by the event rate observed in the past for each user. The event rate remains constant for each user throughout the simulation. In all three models, the probabilities of choosing each event type, of selecting a repository, and of selecting a repository and an event as a pair are determined by frequencies computed from historical training data.


These stationary probabilistic agent-based simulation models compute probabilities for each user agent individually and thus capture individual characteristics of each user, creating a high resolution simulation. The approach is computationally simple enough to be scalable to millions of agents and repositories. Generally, this modeling approach is justified if 
users' future behavior tends to be similar to their past behavior.
A limitation of these models is that they only predict users' interactions with repositories they have interacted in the past. That means that these models should be augmented with additional behavior rules to introduce previously unobserved user-repository interactions. The preferential attachment model is an example where such rules were introduced.

\subsection{Link prediction through embedding}
\label{sec:embedding}
  
 One way of simulating user-repository interactions in GitHub is by predicting the likelihood that a user will perform an event of a certain type on a repository. We can formulate this problem as a link prediction task, by describing our system as a bipartite network in which each node is either a user or a repository and links in each network are specific events. By predicting such links, we can measure the probability of specific events between any given user-repository pair. Each network is built as follows. Nodes belong to either the group of users $U$ or the group of repositories $R$, and a node $u\in U$ is linked to a node $r\in R$ if the user $u$ performed an event on the repository $r$. We weight each link by computing the number of times the user performs that event. As a result, we generate a total of 12 bipartite networks, one for each event type with the exception of \textit{create} and \textit{delete} events.  We can then represent each of the built networks as a weighted adjacency matrix $\mathbf{A}_e\in\mathbb{R}^{|U|\times |R|}$, where $e$ is an event type. However, this link prediction problem does not take into account new users and repositories that have been added (or removed) by the system.
\vspace{-1em}

  \begin{table}[h]
      \centering
      \caption{\small{Average MAP for links and weights prediction on GitHub event networks.}}
      \scalebox{0.8}{
      \begin{tabular}{c|c|c|c|c|c}
      \hline\hline
      & \textit{Push} & \textit{PullRequest} & \textit{IssueComment} & \textit{Fork} & \textit{Watch}\\
      \hline\hline
       Random & 0.01 & 0.03 & 0.02 & 0.03 & 0.03\\
       LE & 0.13 & 0.30 & 0.33 & 0.14 & 0.07\\
       HOPE & 0.13 & 0.25 & 0.26 & \textbf{0.17} & \textbf{0.10}\\
       GF & \textbf{0.25} & \textbf{0.42} & \textbf{0.48} & 0.16 & 0.06\\
      \hline\hline
      \end{tabular}
      }
      \label{MAP}
  \vspace{-1.5em}
  \end{table}

Given the matrix $\mathbf{A}_e$ for each event type $e$, we compare embedding methods against a random baseline: Graph Factorization (GF), Laplacian Eigenmaps (LE), and Hybrid Orthogonal Projection and Estimation (HOPE). We test performance using the MeanAveragePrecision (MAP), which estimates a model precision for each node and computes the average over all nodes, as follows:
\vspace{-0.5em}
\begin{equation}
\footnotesize
     MAP = \frac{1}{|U|+|R|}\sum^{|U|+|R|}_i \frac{\sum_k Pr@k\mathcal{I}\left\{E_{pred,i}(k)\in E_{obs,i}\right\}}{\lvert\left\{k:E_{pred,i}\in E_{obs,i}\right\}\rvert}\;\;,
 \end{equation}
where, $Pr@k = \frac{\lvert E_{pred,i}\left(1:k\right)\cap E_{obs,i}\rvert}{k}$ is the precision at $k$, and $E_{pred,i}$ and $E_{obs,i}$ are respectively the predicted and observed edges for node $i$.

The results are shown in Table~\ref{MAP}, where we consider the $5$ main actions that users can perform on GitHub: contributions (\textit{push} and \textit{pull}), \textit{issues}, and popularity (\textit{fork} and \textit{watch}). Here, we highlighted the highest performance obtained among the methods in each case. All the methods outperform our random baseline, which predicts links in a random fashion. Moreover, GF performs better than the non-linear models in the majority of cases. This is not true for popularity related events, where we find that HOPE is the best predictor. As discussed in Sec.~\ref{sec:discussion}, \textit{forks} are more stable than \textit{watches}. Thus, HOPE is able to capture non-linearity in the \textit{watch} events better then GF. However, GF is comparable to HOPE in predicting \textit{forks}, and it is also much more scalable. Thus, we decide to select GF and its link prediction to inform our agents. 

\subsection{Bayesian model}\label{sec:bayes}
\begin{figure}[hbt]
\centering
\includegraphics[trim={0 4in 0 0.5in}, width=\columnwidth]{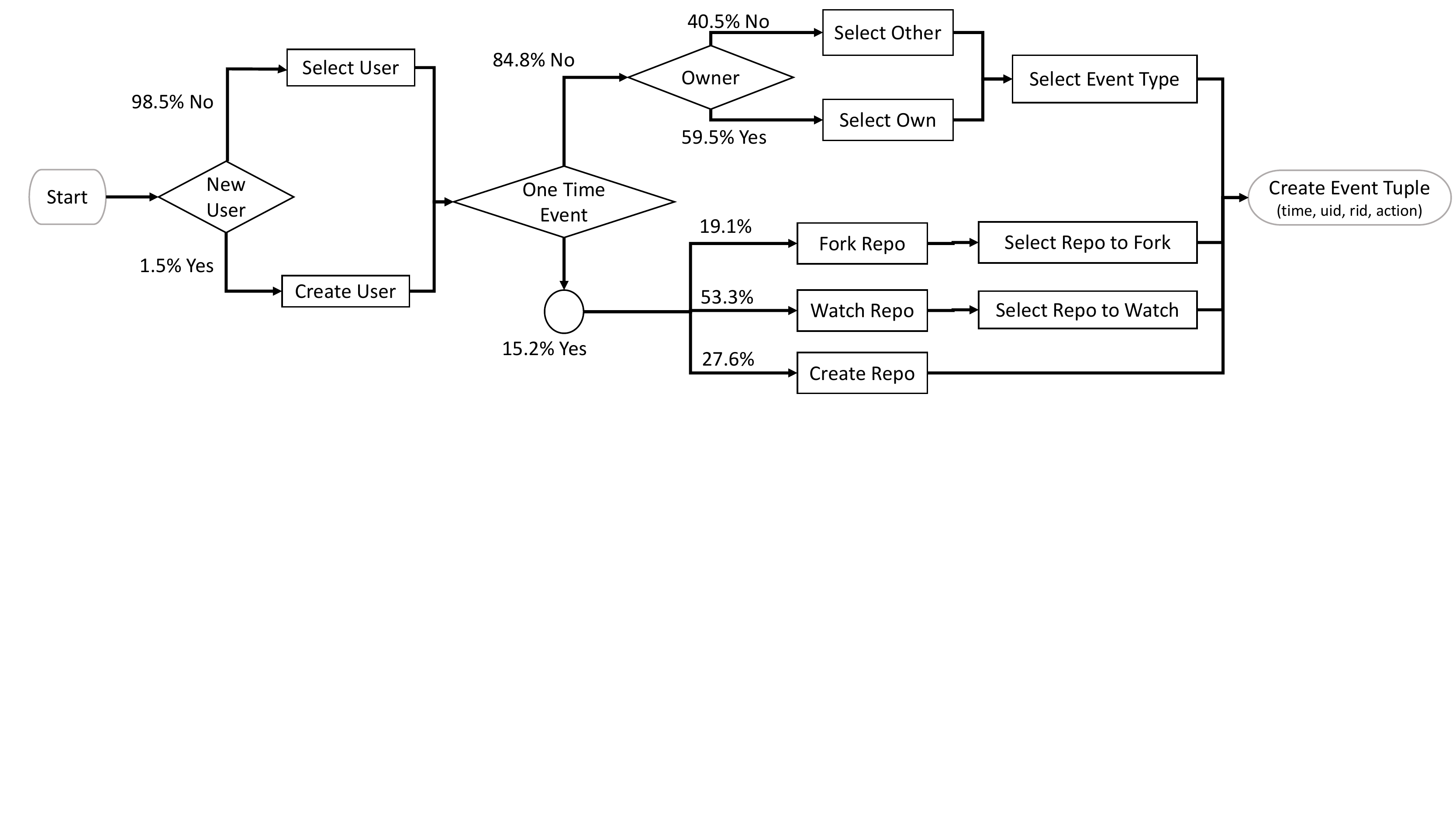}	
\caption{\small{Bayesian model from data inferred frequencies.}}
\label{fig:bayesian-model-diagram}
\vspace{-0.8em}
\end{figure}

The GitHub Challenge can be seen as finding relationships between the three governing entities: users, repositories, and events. We empirically measured the probabilities of these relationships to adjust the posterior probabilities of a generative model. In this section, we present exploratory findings and how they shaped the overall flow of our \textit{Bayesian model}. Figure \ref{fig:bayesian-model-diagram} shows one iteration of the Bayesian model, i.e,. creating a tuple with a user, a repository, and an event type. 
In general terms, the model first chooses between to create a new user or to select an existing one. Then, it decides between a category of events. Finally, it selects a repository and an action to perform.

We investigated the trade-off between recency and history as driving forces to popularity~\cite{Barbosa2015}. The results showed that \textit{less is more} in terms of the amount of data needed to predict users' activity level. For instance, the top-200 most active user rank of a particular month will intersect less with the rank from previous months as we aggregate more data; e.g., while $+75\%$ of the users in the rank remains when the rank is computed only from one month, the intersection drops to 72\%, 59\% and 56\% if we aggregate data from the previous 3, 5, and 7 months, respectively. Recent users, i.e., the ones active in the previous month, represent more than $50\%$ of the active users in the following month, and they are responsible for 4 out of 5 events being generated. 
The user selection implements a rank model~\cite{Fortunato2006} based on user's activity level, with past activity being less weighted using a 30-day half-life decay.
From repository perspective, however, new ones are more likely to be active in the following month than old ones, in absolute numbers (68\% \textit{vs.} 10\%) and in volume of events ($38\%$ \textit{vs.} 5\%). Yet, recent repositories account for almost 3 out of 5 events. 

To address this imbalanced scenario, of increasing newcomers receiving little attention, the model splits between one-time or multiple-time events before choosing the repository.
For instance, the same repository cannot be \textit{created} twice and it is unlikely to be \textit{watched} or \textit{forked} more than once by the same user; conversely, users \textit{push} and \textit{create issues} multiple times on the same repository. The analysis of the event-specific bipartite networks of users and repositories reveals different mechanisms underling them. For instance, in the network formed by \textit{watch} events, the degree distribution of repositories (i.e., the distribution of number of watches per repository), fits a steep power law ($\gamma=1.81$ and $xmin=3$); the \textit{pull request} distribution, however, fits a longer tailed power law ($\gamma=2.54$ and $xmin=291$). 
There are two independent rank models~\cite{Fortunato2006} to select which repository to \textit{watch} or to \textit{fork}. 

Users work more on their repositories (owners) than on other's (contributors). Moreover, they usually behave different depending on these roles.
A owned-repository is selected with probability proportional to the amount of previous activity. Other's repositories are selected using a random walk with small length given by a geometric distribution (mean 2) capturing user's social vicinity. This mechanism assumes a user is more likely to work on repositories he/she already worked with or from previous collaborators. 
Nevertheless, 88\% of the first action of users to a repository (repository discovery) is a one-time event: \textit{watch} (64\%), \textit{fork} (20\%), or \textit{create} (4\%).
Hence, for multiple-time events the model first decides between working on a user's own repositories or not. Although \textit{pushes} are the most prevalent action regardless repository's ownership, some events seemed to be over/under represented depending on ownership. 
For instance, \textit{issue comment}, \textit{commit comment}, and \textit{watch} are more likely to be performed by contributors on others' repositories.


\subsection{Modeling new users and repositories} \label{sec:s3d}
GitHub grows quickly. Almost 50\% of all user accounts that were active in 2017 were created in the same year and a significant fraction of repositories are also new, as mentioned in Section~\ref{sec:bayes}. Such a growth rate yields a system with high number of new active agents. Previously described stationary probabilistic models do not fully address the behaviour of the new agents, since they have no history of interactions. Here, we describe an extension of the models that focuses on predicting the actions of the new agents and newly created relations. Without a historical record of events between user and a repository, the most informative piece of data is missing. Still, both of the entities can be described by a set of their native features, such as: \textit{age} or \textit{number of followers}. We compiled a total list of 124 features extracted for a sample of user-repository pairs. Features include various statistics on user and repository activities for all event types. 

\begin{figure}
    \begin{tabular}{p{0.4\textwidth} p{0.6\textwidth}}
    \vspace{0pt}
    \includegraphics[width=\linewidth]{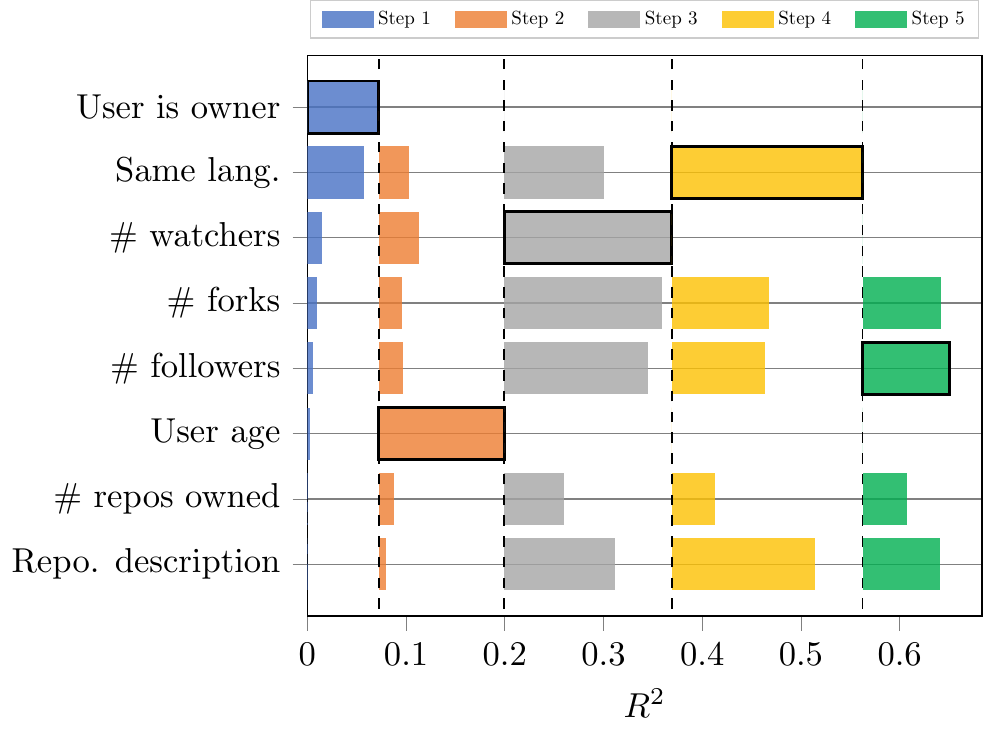} &
    \vspace{0pt}
    \scalebox{0.7}{
\setlength{\tabcolsep}{10pt}
\begin{tabular}{llll} \toprule
Target variable, $\#e$          & 1st feature               & 2nd feature                   & $R^2$      \\
\midrule
\# \textit{Watch}                    & \# watchers & \# repos owned    & 0.306 \\
\# \textit{Push}                     & User is owner  & User age             & 0.199 \\ 
\# \textit{Fork}                     & \# followers       & \# repos owned    & 0.181 \\
\# \textit{Create}                   & User is owner  & Repo. age             & 0.142 \\
\# \textit{PullRequest}              & Same lang.          & \# forks    & 0.109 \\
\# \textit{Delete}                   & Same lang.          & User age             & 0.090 \\
\bottomrule
\end{tabular}
}
\end{tabular}
\vspace*{-1em}
    \caption{\small{On the left, the variation in the target variable \textit{\#PushEvents}, explained by each feature in each consecutive step of the S3D algorithm. On the right, the two most important features to predict the number of events for a new user-repository interaction. The $R^2$ column shows the total gain of the two features.}} 
    \vspace{-1em}
    \label{fig:push_S3D}
\end{figure}

The aim is to build a parsimonious model, able to predict the frequency of a particular event type $e$ performed by user $u$ on a repository $r$, conditioned by non-existing history of interaction between $u$ and $r$. The critical part of model creation is \textit{feature selection}, as we are interested in minimizing the set of features by selecting only the most informative ones. For that purpose, we use Structured Sum of Squares Decomposition (S3D) algorithm~\cite{Fennell2018}. The importance of features is quantified by a measure of variability $R^2$ of outcome variable $Y$ the feature explains. A high value of $R^2$ suggests the strong predictive power. To prevent overfitting, we optimize an inter-model parameter $\lambda$.

For each event type $e$ from the set of event types $E$ we build the predictive model that for a given pair of user and repository $[u,r]$ estimates the number of events $\#e$ performed by the user $u$ on the repository $r$. This way we create 14 different models, one for each event type. With each iteration we select the most informative feature which explains the largest amount of variation in the target variable $\#e$. Each successive feature explains the most of the remaining variation, conditioned on the previous one. The S3D algorithm allows the visual inspection of the selected important features. The iterative process of feature selection is illustrated in Figure~\ref{fig:push_S3D}. To predict the number of \textit{PushEvent}s, a user $u$ is going to perform on repository $r$, the most important feature is the information of the repository ownership. Other features such as \textit{user age}, \textit{\# watchers} or \textit{\# user followers} become important in the following steps. 

The rank of features differ among the models. Still, the most informative features for predicting the actions of a user to a repository, are derived from their mutual relation. Furthermore, many actions depend on the information of the repository ownership 
(Figure~\ref{fig:push_S3D}).
\textit{User is owner} is a boolean variable indicating if the user is the owner of the repository. For some other events, the \textit{Same lang.} variable becomes important. It indicates if the programming language assigned to a repository is the same as the programming language mostly used by a user. 
Even though the $R^2$ can not inform on a model accuracy, it can suggest the potential of a model's prediction performance. 


\section{Candidate Agents and Results}
We developed GitHub user agents that implemented the following models described in the previous section:
(1) the null model,
(2) the probabilistic baseline model,
(3) the probabilistic ground-event model,
(4) the preferential attachment model,
(5) link prediction via embedding (LPE),
and (6) the Bayesian model.

The null model is just a shift of the past data to the future. In this case we used two weeks immediately preceding the test period and returned it, with dates adjusted, as the new prediction. Both baseline model and null model were used to compare relative performance of other agents. The LPE model was the best performing of a number of machine learning-based models that were explored.

The DARPA SocialSim Challenge design had multiple research questions in mind. There were five groups of research questions that covered user engagement, contributions, reputation, influence and popularity. To evaluate fidelity of the simulation a set of metrics was developed by PNNL. These metrics evaluate simulation fidelity on three levels: node-level, community-level and population-level.
Although it is not feasible to discuss all metrics in the context of this paper, we selected the following five to discuss in detail, which cover user and repository metrics on individual, community and population levels.
\\
{\bf user popularity} - top 500 most popular users, measured as the total number of  \textit{watch} and \textit{fork} events on repositories owned by user. Calculated as Rank-Biased Overlap ($RBO$)~\cite{Webber2010}.
\\
{\bf repository popularity} - top 500 popular repositories. Calculated as $RBO$.
\\
{\bf repository event count issues} - the number of  \textit{issue} events by repository. Calculated as $R^2$.
\\
{\bf repository contributors} - the number of daily unique contributors to a repository as a function of time. Calculated as Root Mean Square Error ($RMSE$).
\\
{\bf community contributing users} - percentage of users who perform any contributing (e.g.  \textit{push, pull request} events) actions within the community. 

\begin{figure}[t]
\centering
\includegraphics[trim={1in 0 4.75in 0}, width=0.4\columnwidth]{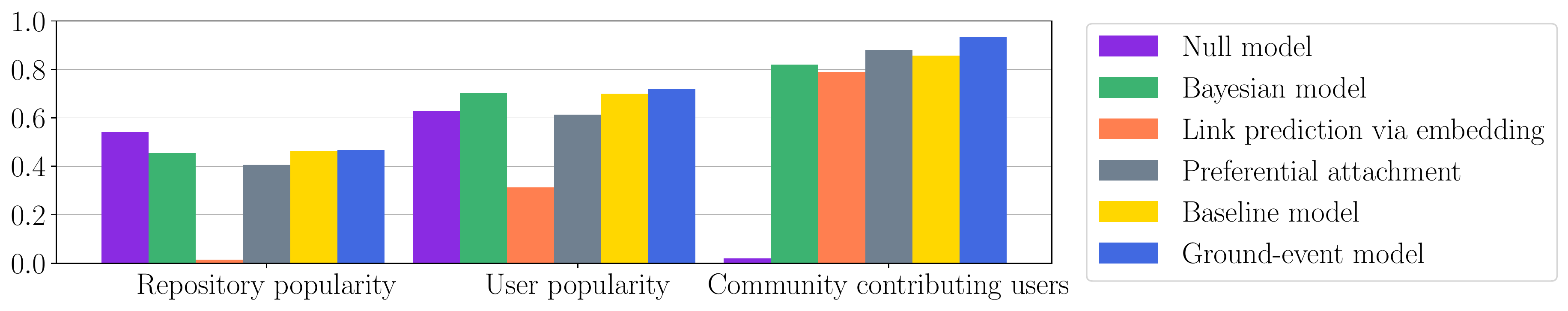}	
\includegraphics[width=0.55\columnwidth,height=0.75in]{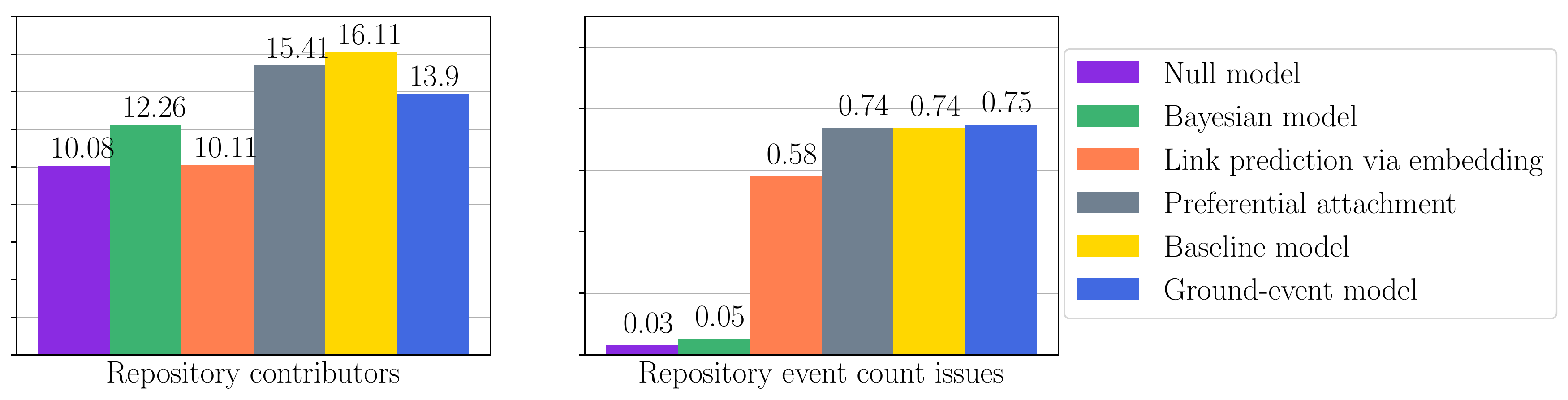}	
\caption{\small{Left: popularity metrics, $RBO$, and community contributing users, higher is better. Center: Repository contributors, $RMSE$, lower is better and right: event issue count, $R^2$, higher is better.}}
\label{fig:popularity}
\vspace{-0.3cm}
\end{figure}

To answer various research questions of the DARPA SocialSim Challenge more than a dozen metrics were used to evaluate our simulation results. 
Figure \ref{fig:popularity} on the left shows evaluation results for two bounded metrics: repository popularity, user popularity. All metrics are scaled to the [0,1] interval, higher is better. All simulation models except the link prediction via embedding model showed comparable results. Note that the popularity metric is not calculated for users and repositories that were created during the test period.

Predicting the 500 most popular users and repositories is difficult with a constant rate of events, due to high month to month turn out. On average about 30\% of repositories in the top 500 change every month. Therefore, we can see that even the null model works better with respect to repository popularity metric.

The {\sc LPE} model uses generalized rules to predict the user's action, which some cases leads to long-tail distributions of probabilities of repository selection. That process introduces noise and the model was also computationally intensive because it requires reconstruction of the \textit{complete user x repository matrix}. To optimize memory usage when repository selection probabilities were computed we truncated long tails (repositories with small probabilities). We used a threshold of 100 repositories per user. The ground-event model showed slightly better results on popularity metrics than the baseline model and the preferential attachment model. All three models use the same approach to compute rate of users' activity but different ways to choose repositories. The preferential attachment model did not leverage much from explicitly modeling fork events and selecting repositories that also popular across user's neighbors.

The {\sc LPE} model generally showed poor results on repository-centered metrics.
All models except the null model had a high percentage (more than 0.75) of community contributing users. The only two models that performed better than the baseline were the preferential attachment model and ground-event model.

Figure  
\ref{fig:popularity}
shows repository contributors - the number of daily unique contributors to a repository as a function of time. It is calculated as $RMSE$, lower is better. The null model and the link prediction model showed the lowest values here. 
\ignore{
\begin{figure}[t]
\centering
\includegraphics[width=0.7\columnwidth,height=1in]{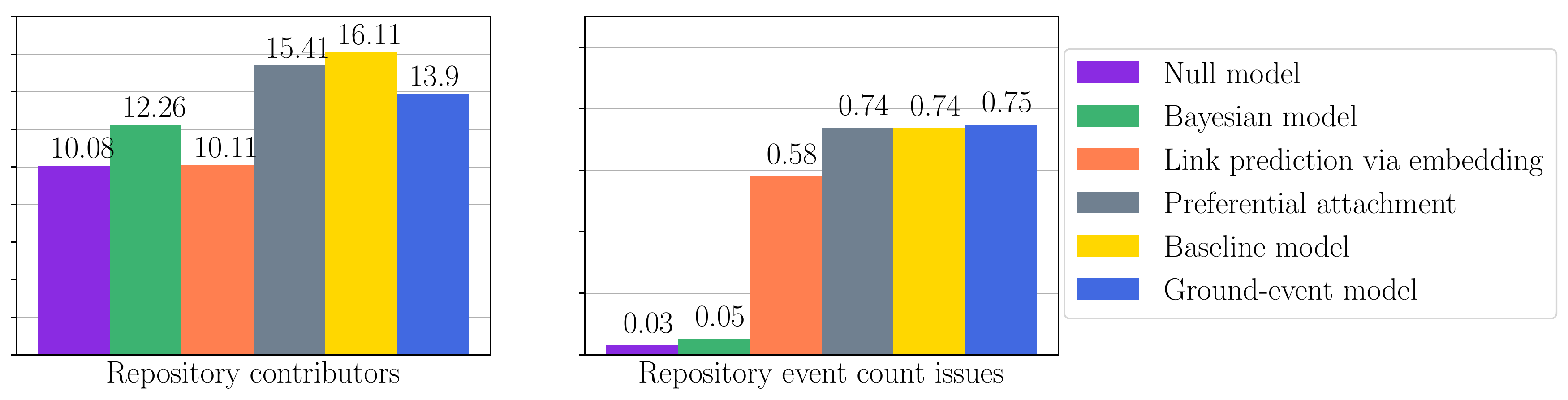}	
\caption{\small{Repository contributors, $RMSE$, lower is better (on the left) and event issue count, $R^2$, higher is better (on the right).}}
\label{fig:contributors_issues}
\vspace{-0.3cm}
\end{figure}
}
Figure  
\ref{fig:popularity}
also shows repository event count issues - the number of issue events by repository. It is calculated as $R^2$, higher is better. Both the null model and the bayesian model show much lower value compare to all other values. The stationary models capture user-repository interactions with better precision because they rely only on training data specific to that user-repository pair.

\section{Related work}
\label{sec:related}


Many groups have addressed large-scale agent simulations over the years. Tumer et al. consider simulations for air traffic flow management, handling more than 60,000 flights~\cite{Tumer2007DAA13291251329434}. Distributed simulations have been developed
~\cite{cosenza2011distributed,vsivslak2009distributed,collier2013parallel}, 
with agents typically partitioned by geographic location and migrated between compute hosts as they move in the simulated space. Noda describes a social simulation of tens of thousands of agents to help explore policies for urban traffic and disaster response~\cite{noda18}. Repast has been used to simulate up to $10^9$ agents in a cascade simulation on supercomputers~\cite{collier2013parallel}. To our knowledge, the simulations in response to the SocialSim challenge are the first systematic experiments involving millions to tens of millions of agents with individual complex behavior.

Researchers have given an overview of the GitHub platform and user/repository statistics
~\cite{Gousios:2013:GDT:2487085.2487132,6224294,Dabbish2012SocialCI}. 
The amount of available data on GitHub 
allows observing group behaviour through the collaboration network structure~\cite{Thung:2013:NSS:2495256.2495709} and studying the interaction patterns inside teams on public GitHub 
repositories
~\cite{klug2016understanding,Sornette2014,Lima2014CodingTA}. 
Another line of research focuses on the success and popularity of open-source projects. Successful and popular repositories usually have consistent documentation ~\cite{Aggarwal:2014:CPD:2597073.2597120,Zhu:2014:PFU:2652524.2652564}, various growth patterns~\cite{Borges2016PredictingTP}, and popular programming languages~\cite{6649842}.

\section{Discussion}
\label{sec:discussion}

We demonstrated novel agents built using six different learning principles to predict the future behavior of GitHub based on training data. The agents were tested in the same simulation environment with the same implementation of GitHub actions. Across a broad array of prediction metrics, no single approach dominated the others. The Bayesian model performed well on user popularity while the stationary per-agent action distributions performed better on predicting contributors and event counts. One interesting constant across all the models was that, since overall behavior is constantly changing, it was detrimental to use all available training data in building the agents. Instead, one month of data proved optimal across most of the agents, although this precise number is no doubt dependent on the social network in question.


The main contribution of this work is to develop a framework for massive-scale simulations in which agents embodying very different ideas about decision making and data use can be directly compared. 
Our approach is general, and has recently been applied to the Reddit and Twitter social networks.
We are also considering ways to combine these agent models, both intra-agent, combining some of the best features of different approaches in a single agent, and inter-agent, with simulations with more than one type of agent.

Predicting new \textit{user-repository} pairs requires the development of more general models that will be used to predict new pairings among different types of entities. While developing the probabilistic models, we were constantly faced with a particular chicken-egg problem: which pair to choose the first from the \textit{user-repository-event} triad. The models implemented in the simulation make use of previously established links between users and repositories. We are developing a model to predict the number of events performed by a user both on old and new repositories. For a given user and type of event, the model will be able to identify a small subset of candidate repositories for the event, regardless of the previous interaction between the observed user and their repositories.

Addressing the new users and repositories is yet another challenge. Modeling the behaviour of entities never seen before can not rely on the historical records. Almost 20\% of events in recent months have been performed by new users. Both new and old users continuously create new repositories. Even with such a high growth rate, GitHub is considered to be relatively slow-paced compared to some other systems, such as Twitter or Reddit. One way to address the newly created entities is to observe some latent features inferred from the user who created it or from the characteristics of the entity itself, avoiding the reference to the past. Therefore, we are continuing to develop the set of models explained in~\ref{sec:s3d}. 


{\bf Acknowledgments:}
The authors are grateful to the Defense Advanced Research Projects Agency  (DARPA), contract W911NF-17-C-0094, for their support.

\bibliographystyle{splncs04}
\bibliography{sample-bibliography,diogo,agent-refs}

\begin{thebibliography}{10}
\providecommand{\url}[1]{\texttt{#1}}
\providecommand{\urlprefix}{URL }
\providecommand{\doi}[1]{https://doi.org/#1}

\bibitem{Aggarwal:2014:CPD:2597073.2597120}
Aggarwal, K., Hindle, A., Stroulia, E.: Co-evolution of project documentation
  and popularity within github. In: Mining Software Repositories. MSR (2014)

\bibitem{Barbosa2015}
Barbosa, H., de~Lima-Neto, F.B., Evsukoff, A., Menezes, R.: {The effect of
  recency to human mobility}. EPJ Data Science  \textbf{4}(1),  1--14 (dec
  2015)

\bibitem{6649842}
Bissyandé, T.F., Thung, F., Lo, D., Jiang, L., Réveillère, L.: Popularity,
  interoperability, and impact of programming languages in 100,000 open source
  projects. In: 2013 IEEE 37th Annual Computer Software and Applications
  Conference (2013)

\bibitem{blythe12}
Blythe, J.: A dual-process cognitive model for testing resilient control
  systems. In: 5th International Symposium on Resilient Control Systems. pp.
  8--12 (Aug 2012)

\bibitem{blythe-tregubov18}
Blythe, J., Tregubov, A.: Farm: Architecture for distributedagent-based social
  simulations. In: IJCAI/AAMAS Workshop on Massively Multi-Agent Systems (2018)

\bibitem{Borges2016PredictingTP}
Borges, H., Hora, A.C., Valente, M.T.: Predicting the popularity of github
  repositories. In: PROMISE (2016)

\bibitem{collier2013parallel}
Collier, N., North, M.: Parallel agent-based simulation with repast for high
  performance computing. Simulation  \textbf{89}(10),  1215--1235 (2013)

\bibitem{cosenza2011distributed}
Cosenza, B., Cordasco, G., De~Chiara, R., Scarano, V.: Distributed load
  balancing for parallel agent-based simulations. In: Parallel, Distributed and
  Network-Based Processing (PDP), 2011 19th Euromicro International Conference
  on. IEEE (2011)

\bibitem{Dabbish2012SocialCI}
Dabbish, L.A., Stuart, H.C., Tsay, J., Herbsleb, J.D.: Social coding in github:
  transparency and collaboration in an open software repository. In: CSCW
  (2012)

\bibitem{Fennell2018}
Fennell, P., Zuo, Z., Lerman, K.: {Predicting and Explaining Behavioral Data
  with Structured Feature Space Decomposition}  (2018),
  \url{https://arxiv.org/abs/1810.09841}

\bibitem{Fortunato2006}
Fortunato, S., Flammini, A., Menczer, F.: {Scale-free network growth by
  ranking}. Physical Review Letters  \textbf{96}(21),  218701 (may 2006)

\bibitem{6224294}
Gousios, G., Spinellis, D.: Ghtorrent: Github's data from a firehose. In: 2012
  9th IEEE Working Conference on Mining Software Repositories (MSR) (June 2012)

\bibitem{Gousios:2013:GDT:2487085.2487132}
Gousios, G.: The ghtorent dataset and tool suite. In: Proceedings of the 10th
  Working Conference on Mining Software Repositories. MSR '13, IEEE Press
  (2013)

\bibitem{klug2016understanding}
Klug, M., Bagrow, J.P.: Understanding the group dynamics and success of teams.
  Royal Society open science  \textbf{3}(4),  160007 (2016)

\bibitem{Lima2014CodingTA}
Lima, A., Rossi, L., Musolesi, M.: Coding together at scale: Github as a
  collaborative social network. CoRR  \textbf{abs/1407.2535} (2014)

\bibitem{noda18}
Noda, I.: Multi-agent social simulation for social service design. In:
  IJCAI/AAMAS Workshop on Massively Multi-Agent Systems (2018)

\bibitem{vsivslak2009distributed}
{\v{S}}i{\v{s}}l{\'a}k, D., Volf, P., Jakob, M., P{\v{e}}chou{\v{c}}ek, M.:
  Distributed platform for large-scale agent-based simulations. In: Agents for
  Games and Simulations. Springer (2009)

\bibitem{Sornette2014}
Sornette, D., Maillart, T., Ghezzi, G.: {How Much Is the Whole Really More than
  the Sum of Its Parts? 1 + 1 = 2.5: Superlinear Productivity in Collective
  Group Actions}. PLoS ONE  \textbf{9}(8),  e103023 (aug 2014)

\bibitem{Thung:2013:NSS:2495256.2495709}
Thung, F., Bissyande, T.F., Lo, D., Jiang, L.: Network structure of social
  coding in github. In: Software Maintenance and Reengineering. CSMR (2013)

\bibitem{Tumer2007DAA13291251329434}
Tumer, K., Agogino, A.: Distributed agent-based air traffic flow management.
  In: Autonomous Agents and Multiagent Systems. AAMAS '07, ACM (2007)

\bibitem{Webber2010}
Webber, W., Moffat, A., Zobel, J.: {A similarity measure for indefinite
  rankings}. ACM Transactions on Information Systems  \textbf{28}(4),  1--38
  (nov 2010)

\bibitem{Zhu:2014:PFU:2652524.2652564}
Zhu, J., Zhou, M., Mockus, A.: Patterns of folder use and project popularity: A
  case study of github repositories. In: Proc. 8th ACM/IEEE International
  Symposium on Empirical Software Engineering and Measurement. ESEM '14, ACM
  (2014)

\end{thebibliography}

\end{document}